\documentclass{emulateapj}
\usepackage{natbib}
\usepackage{multirow}
\shorttitle{Kinematic Modeling of NGC 4013}
\shortauthors{Zschaechner \& Rand}

\begin{document}

\title{The {\sc H\,i} Kinematics of NGC 4013: a Steep and Radially Shallowing Extra-planar Rotational Lag} 
\author{\sc Laura K. Zschaechner\altaffilmark{1,2} \& Richard J. Rand\altaffilmark{2}}
\altaffiltext{1}{Max Planck Institute f\"{u}r Astronomie - K\"{o}nigstuhl 17, 69117 Heidelberg - Germany; zschaechner@mpia.de}
\altaffiltext{2}{Department of Physics and Astronomy, University of New Mexico, 1919 Lomas Blvd NE, Albuquerque, New Mexico 87131 - USA; rjr@phys.unm.edu}
\slugcomment{Accepted to Astrophysical Journal on June 16, 2015}
\begin{abstract}

\par
     NGC 4013 is a distinctly warped galaxy with evidence of disk-halo activity.  Through deep {\sc H\,i} observations and modeling we confirm that the {\sc H\,i} disk is thin (central exponential scale height of with an upper limit of 4" or 280 pc), but flaring.  We detect a vertical gradient in rotation velocity (lag), which shallows radially from a value of  $-$35$^{+7}_{-28}$ km s$^{-1}$ kpc$^{-1}$ at 1.4' (5.8 kpc), to a value of zero near R$_{25}$ (11.2 kpc).  Over much of this radial range, the lag is relatively steep.  Both the steepness and the radial shallowing are consistent with recent determinations for a number of edge-ons, which have been difficult to explain. We briefly consider the lag measured in NGC 4013 in the context of this larger sample and theoretical models, further illuminating disk-halo flows.
\end{abstract}

\section{Introduction}
\par
    To fully comprehend star formation (SF) processes, we must understand the properties and origins of the gas that fuels them.  If galaxies are to avoid exhausting their star forming fuel in a short time compared to their age, then there must exist some means of gaining material, possibly through stellar mass loss as suggested by \citet{2011ApJ...734...48L}, or through the accretion of primordial neutral and ionized gas.  Thus, extra-planar layers are of great interest as they are the interface between the main disk and the intergalactic medium (IGM) and all material accreted onto the midplane to fuel SF must first pass through them.   

\par
    With the exception of neutral hydrogen {\sc H\,i}, a connection has been made between the presence of extra-planar components and SF both on local and global scales (e.g.\ \citealt{2012ARA&A..50..491P} and references therein).  In particular, connections have been made between SF and hot gas (e.g.\ \citealt{2006A&A...448...43T}, \citealt{2013ApJ...762...12H}, \citealt{2013MNRAS.435.3071L}), relativistic particles (e.g.\ \citealt{1999AJ....117.2102I}), dust (e.g.\ \citealt{1999AJ....117.2077H}), and diffuse ionized gas (DIG) (e.g.\ \citealt{1996ApJ...462..712R}; \citealt{2003A&A...406..493R}). Extra-planar {\sc H\,i} has been shown to have some connection to disk-halo flows within some galaxies (e.g.\ \citet{1994ApJ...429..618I} and \citet{2005A&A...431...65B}), but evidence for {\sc H\,i} accretion [e.g.\ HVCs in the Milky Way \citep{1997ARA&A..35..217W},  counterrotating clouds in NGC 891 seen by \citealt{2007AJ....134.1019O}] exists, leaving the relative contributions of each process ambiguous until further study.

\par
   A powerful method to determine the origins of extra-planar layers is to study their kinematics.  In particular, decreases in rotation speed with height above the midplane (lags) have come to the forefront in recent years (e.g.\ \citealt{2007AJ....134.1019O}, \citealt{2007ApJ...663..933H}) leading to various models attempting to explain them in terms of disk-halo flows or an interaction between disk-halo cycling gas and accreting gas.  Purely ballistic models of disk halo flow (e.g.\ \citealt{2002ApJ...578...98C}, \citealt{2006A&A...446...61B}, \citealt{2006MNRAS.366..449F}) consider only gravitational effects:   Material is ejected from the midplane in disk-halo flows and then moves outward radially due to its being farther from the gravitational potential.  To conserve angular momentum, the rotational velocity of the material decreases, creating a lag with height above the midplane.  A clear shortcoming of such models is that they greatly under-predict lag magnitudes by up to an order of magnitude \citep{2007ApJ...663..933H}.  Thus, additional processes may be at work such as pressure gradients or magnetic tension (Benjamin 2002, 2012).  However, simulations involving these mechanisms have yet to be developed in earnest, with future progress made possible through observational constraints, in particular those involving lags.

\par
   Pressure gradients and magnetic tension aside, some theoretical simulations have gone beyond purely ballistic effects and disk-halo flows considered in isolation.  Those presented by \citet{2011MNRAS.415.1534M} involve momentum and heat exchanges occurring in disk-halo flows.  When cool clouds are launched from the disk into a hot, initially static corona built up by accretion, gas is stripped from the clouds, which streams in the cloud's wake, mixing with the hot coronal gas.  The mixed gas forms additional {\sc H\,i} clouds before falling to the midplane.  This mixing results in substantially more efficient cooling of the hot coronal gas than is described in the earlier, related model of \citet{2008MNRAS.386..935F}, in which hot coronal gas experienced cooling times $\sim$100 times greater than the expected flow time of galactic fountain clouds, thus resulting in most of the hot gas remaining in the halo.  \citet{2011MNRAS.415.1534M} demonstrate that this process would contribute approximately $-$7 km s$^{-1}$ kpc$^{-1}$ to the deceleration of clouds, amounting to 50$\%$ of the total $-$15 km s$^{-1}$ kpc$^{-1}$ observed in NGC 891.  If one also considers ballistic effects, this momentum exchange would nearly eliminate discrepancies between purely ballistic models and observed lags in NGC 891 and the Milky Way.  However, lags have been shown to differ substantially among galaxies (e.g.\ \citealt{2007ApJ...663..933H}, \citealt{2008A&ARv..15..189S}, \citealt{2012ApJ...760...37Z}, \citealt{2013MNRAS.434.2069K}).  Thus, additional galaxies must be observed and modeled to comparable sensitivity and detail.  The Westerbork Hydrogen Accretion in LOcal GAlaxieS (HALOGAS) survey \citep{2011A&A...526A.118H} has sought to remedy this situation through extensive {\sc H\,i} observations of a carefully selected sample of nearby spiral galaxies.  

\par
     Not only are lag magnitudes of interest, but variations in lag magnitude with radius are expected based on conservation of angular momentum. From ballistic considerations only, in a disk-halo flow, material at smaller radii experiences larger relative changes in the gravitational potential when ejected from the midplane to a given height (and thus a greater lag) than material at large radii.  \citet{2007AJ....134.1019O}, Zschaechner et al. (2011, 2012), and \citet{2011MNRAS.414.3444K} measure a shallowing of the lags in several nearby galaxies, with a current summary given in \citet{2015ApJ...799...61Z}.  A general trend noted in that paper is that the lags begin to shallow near a radius of 0.5R$_{25}$, and reach their shallowest point near R$_{25}$ itself.  Considering the mismatch between observed lag magnitudes and ballistic models, it is likely that the origin of the shallowing is more complicated as well.

\par
     This work builds upon the HALOGAS sample and models, as well as supplemental galaxies presented in \citet{2015ApJ...799...61Z}.  To this existing collection of modeled galaxies, we add NGC 4013, a nearby, well-studied, edge-on spiral galaxy with multiple extended extra-planar components.

\subsection{NGC 4013}
\par
   NGC 4013 is a member of the NGC 4151 group containing 16 galaxies, but is itself rather isolated \citep{2000ApJ...543..178G}. Arguably, its most noteworthy feature is its substantial warp originally seen in {\sc H\,i} by \citet{1987Natur.328..401B}.  Here we present detailed {\sc H\,i} observations and tilted-ring models of this galaxy.  A brief summary of NGC 4013's observational properties is provided in Table~\ref{n4013tbl-1}.  Before presenting our work, we will review some of the key findings throughout the literature concerning NGC 4013.

\par
   \citet{1996ApJ...462..712R} presents an analysis of the DIG in nine nearby, edge-on spiral galaxies, including NGC 4013.  In that work, four extra-planar filaments are found to extend 2.5 kpc from the disk, defining a nearly ``H" or ``X" shape.  These filaments could be related to outflows fueled by SN activity near the center.  The extra-planar DIG is not nearly as prominent as in NGC 891 or NGC 5775, which have higher SFRs.  The presence of extended extra-planar DIG is confirmed by \citet{2003ApJ...592...79M}, who also see indications of gradients in rotation speed with height above the midplane.  \citet{2013AJ....145...62R} observe the DIG, but also find extra-planar dust up to approximately 2 kpc above the disk.  The dust is highly structured and filamentary - far more so than the DIG.  They conclude that direct evidence for a connection between the extra-planar DIG and extra-planar dust is lacking, suggesting the gas associated with the dust is either atomic or molecular.

\par
    \citet{2013A&A...556A..54V} observe the dust in seven nearby, edge-on spiral galaxies using Herschel.  According to their analysis, NGC 4013 is the only galaxy within their sample to display substantial extra-planar dust.  

\par 
     All of these results suggest disk-halo cycling at a level above those of galaxies such as NGC 4244 and NGC 4565, but lower than that of NGC 891.

\par
   \citet{1999A&A...343..740G} present CO observations of NGC 4013, emphasizing its box-shaped bulge that is likely due to a bar.  They note extra-planar extensions showing some connection to the extra-planar DIG described by \citet{1996ApJ...462..712R}.  Additionally, they find a 130 km s$^{-1}$ spread in velocities near the center, which they note has no corresponding {\sc H\,i} feature.

\par
    \citet{2009ApJ...692..955M} discovered a tidal stellar stream extending 3 kpc above the plane of the disk in NGC 4013, possibly indicating a minor merger has taken place approximately 3 Gyr ago.  Such a scenario was previously unexpected, as NGC 4013 appears to be relatively isolated and undisturbed.

\par
   \citet{2014AAS...22324619R} observe an extra-planar {\sc H\,ii} region in NGC 4013 that is 850 pc above the midplane and has a substantially lower metallicity than the main disk.  They interpret this as a mixing of material in the gas that formed the associated stars.

 
\par
    Perhaps most relevant to the work presented here, {\sc H\,i} observations and modeling were presented in \citet{1995A&A...295..605B}.  Those observations were done using the WSRT, achieving a spatial resolution of 19"$\times$13", velocity resolution of 33.3 km s$^{-1}$, and a single channel rms noise of 0.3 mJy beam$^{-1}$.  As will be shown in the next section, our data improve upon these numbers.

\par
   \citep{1996A&A...306..345B} presented tilted-ring models of NGC 4013 using the aforementioned WSRT data.  Those models include a central inclination of 90$^\circ$, and a central position angle of 66$^\circ$.  A substantial warp is also modeled. The rotation curve peaks at 195 km s$^{-1}$ and decreases to 165 km s$^{-1}$ at large radii.  Naturally, the models presented in this paper will also constrain these parameters, which we derive independently but compare in $\S$~\ref{4013discussion}.  Additionally, the models presented here focus largely on the extra-planar dynamics, which were not considered in \citep{1996A&A...306..345B}.

\begin{deluxetable*}{lrr}
\tabletypesize{\scriptsize}
\tablecaption{NGC 4013 Parameters\label{n4013tbl-1}}
\tablewidth{0pt}
\tablehead
{
\colhead{Parameter} &
\colhead{Value}&
\colhead{Reference}\\
}
\startdata
\phd Distance (Mpc) &14.6\tablenotemark{a} &Nasa Extragalactic Database\\ 
\phd Systemic velocity (km s$^{-1}$)&835 &Bottema (1995)\\
\phd Inclination &90$\,^{\circ}$ &This work\\
\phd Morphological Type &SBc& \citet{1988JRASC..82..305T} \\
\phd SFR ($M_{\odot}yr^{-1}$) &0.48&Weigert et al. (2015, \textit{submitted to AJ})\\
\phd Kinematic Center $\alpha$ (J2000.0) & 11h 58m 31.6s &This work\\
\phd Kinematic Center $\delta$ (J2000.0) & 43d 56m 52.2s &This work\\
\phd D$_{25}$ (kpc) &22.3& \citet{1991trcb.book.....D}\\
\phd Total Atomic Gas Mass ($10^{9}M_{\odot}$) &2.1 \tablenotemark{c}&This work\\
\enddata
\tablenotetext{a}{Distance is the median value of distances found on the NED database, excluding those obtained using the Tully-Fisher relation to be consistent with HALOGAS.}
\tablenotetext{b}{Includes neutral He via a multiplying factor of 1.36.  Value obtained using single dish data at our adopted distance is 2.4$\times10^{9}M_{\odot}$ \citep{2005ApJS..160..149S}.  WSRT value obtained by \citet{1995A&A...295..605B} using our adopted distance is 2.3$\times10^{9}M_{\odot}$. }
\end{deluxetable*}

\section{Observations \& Data Reduction}

\par
   Observations of NGC 4013 were completed with the VLA in B and C configurations.  Data reduction was performed in AIPS using standard spectral line methods.  Self-calibration was performed after combining all tracks, but before continuum subtraction, and led to substantial improvement.  Although the velocity resolution of these data is 6.7 km s$^{-1}$, due to incomplete (not all of the awarded time was observed) observations, we average three channels to obtain our desired rms noise level for the final cube.  A variety of weighting schemes were used to make cubes, with the final, full resolution cleaned cube using Briggs weighting and a robust parameter of 3 after attempting a range of robust parameters to find the optimal balance between resolution and sensitivity.  A low spatial resolution cube (used only for the lowest contour in Figure~\ref{totalHIdata4013}) uses a robust parameter of 2 with an outer \textit{uv} taper of 7 kilo-lambda.  Observation dates and cube parameters are given in Table~\ref{4013tbl-2}.  The primary beam correction was performed in AIPS.

\begin{deluxetable}{lr}
\tabletypesize{\scriptsize}
\tablecaption{Observational and Instrumental Parameters  \label{4013tbl-2}}
\tablewidth{0pt}
\tablehead
{
\colhead{Parameter} &
\colhead{Value}\\
}
\startdata
\phd Observation Dates $-$ C Config.& 2010 Nov 16\\
\phd &2010 Nov 20\\
\phd &2010 Nov 28\\
\phd B Configuration &2011 Mar 21-23\\
\phd &2011 Apr 02\\
\phd &2011 Apr 22\\
\phd &2011 May 02\\
\phd &2011 May 04\\
\phd &2011 May 09\\
\phd Total Time on source - C Config. (hours)&9\tablenotemark{a}\\
\phd Total Time on source - B Config. (hours)&9.5\tablenotemark{b}\\
\phd Pointing Center &11h 58m 31.30s\\
\phd &43d 56m 48.00s\\ 
\phd Number of channels & 256\\
\phd Velocity Resolution &20.1 km s$^{-1}$\tablenotemark{c}\\
\phd Primary Cube Beam Size &12.7$\times$9.91"\\
\phd &900$\times$700 pc\\
\phd PA &4.92$^{\circ}$\\
\phd RMS Noise $-$ 1$\sigma$ 1 chan. &0.18 mJy beam$^{-1}$\\
\phd Secondary Cube Beam Size &23.4$\times$19.7"\\
\phd &1700$\times$1400 pc\\
\phd PA &84.37$^{\circ}$\\
\phd RMS Noise $-$ 1$\sigma$ 1 chan. &0.23 mJy beam$^{-1}$\\

\enddata
\tablenotetext{a}{Although nine total hours were observed, approximately three of these hours had substantial RFI and were flagged heavily.}
\tablenotetext{b}{Approximately two hours of these data were omitted due to poor quality, resulting in 7.5 hours used in the final cubes.}
\tablenotetext{c}{Averaged from 6.7 km s$^{-1}$ to 20.1 km s$^{-1}$ for modeling in order to improve SNR.}
\end{deluxetable}

\section{The Data}

\begin{figure}
\centering
\includegraphics[width = 80mm]{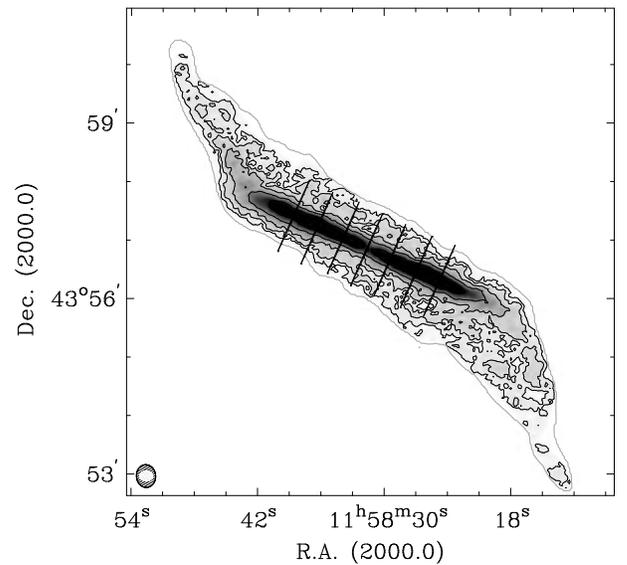}
\caption[A Zeroth-moment Map of NGC 4013]{\scriptsize\textit{A zeroth-moment map of NGC 4013.  The full resolution cube is represented in black, while an outline of the smoothed cube is in gray.  Contour levels for the full resolution cube begin at 2.6$\times$10$^{20}$ cm$^{-2}$ and increase by factors of two.  The gray contour is at 9.5$\times$10$^{19}$ cm$^{-2}$.  Black lines represent the slice locations of bv diagrams in Figure~\ref{4013bvapproaching}.  The beams are shown in the lower left-hand corner (white is full resolution, black is the smoothed cube).  Note the prominent warp.  Also note the apparent thickness of the warped regions, possibly due to a flare, thick disk, or a warp component along the line of sight.} \label{totalHIdata4013}}
\end{figure}

\par
    Immediately clear from the zeroth-moment map (Figure~\ref{totalHIdata4013}) is the prominent, nearly symmetric warp in NGC 4013.  Hints of a line of sight warp component, flare, or possibly a thickened {\sc H\,i} layer may also be seen extending to an \textit{apparent} height of $\sim$0.7' (3 kpc).  Thickening of the disk largely to the southwest strongly suggests evidence for a line of sight warp component, although the other possibilities (or a combination of them) cannot yet be ruled out. It is unlikely that the warp is oriented across the line of sight - its more probable that the warp is indeed comprised of both across and along line of sight components.  Further investigation through modeling ($\S$~\ref{models}) is required to determine which of these components are actually present in NGC 4013.   The final constraints on these components will be given in $\S$~\ref{4013discussion}.  

\par
\begin{figure*}
\centering
\includegraphics[angle=270, width =160mm]{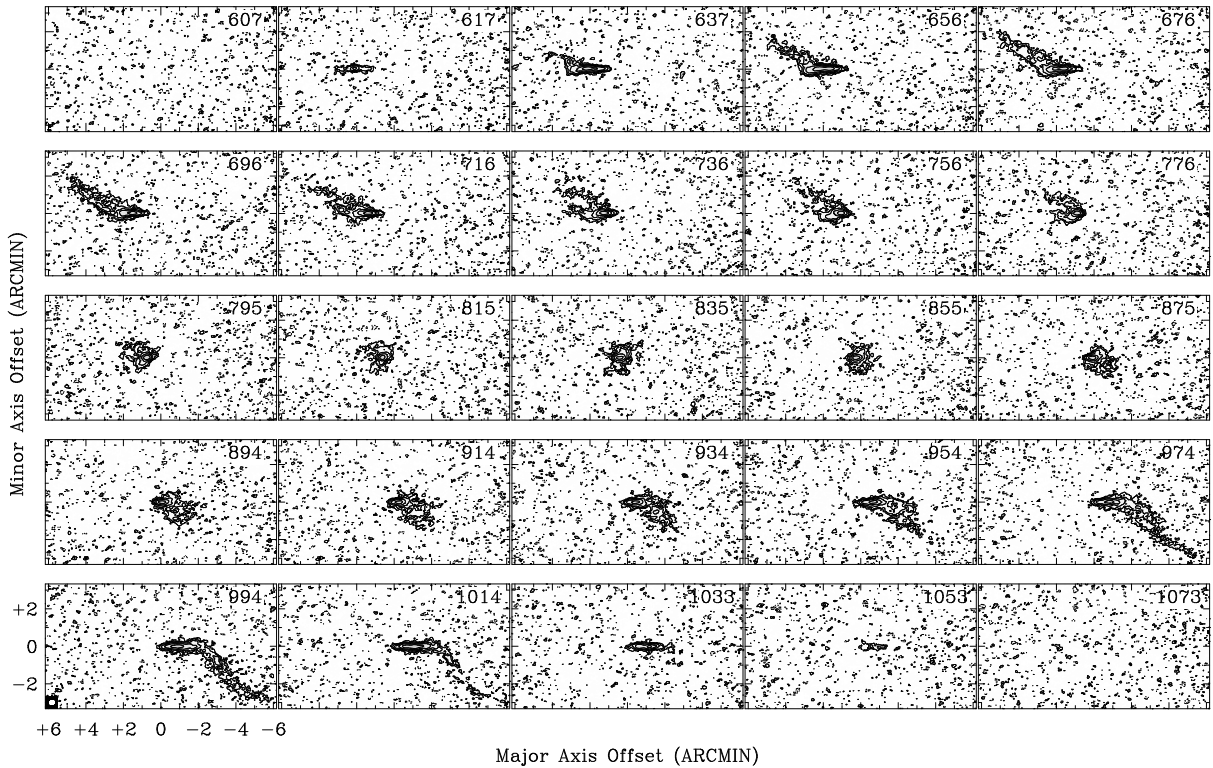}
\caption[Channel Maps of NGC 4013]{\scriptsize\textit{Channel maps of NGC 4013 show the substantial warp.  Velocities in km s$^{-1}$ are given in each panel.  Contours begin at 2$\sigma$ (0.36 mJy beam$^{-1}$, or 6.4$\times$10$^{19}$ cm$^{-2}$) and increase by factors of two.  The cube is rotatied 66$^{\circ}$.}} \label{channelmapsymmetric4013}
\end{figure*}

   Channel maps (Figure~\ref{channelmapsymmetric4013}) again show the prominent warp in NGC 4013.  Evidence for a line of sight warp component may be seen by the splitting in channels corresponding to 736 km s$^{-1}$ and 756 km s$^{-1}$.  It is clear that the disk is very thin near the center.  Note that more than half of the total radial extent is clearly affected by the warp.  The component of the warp across the line of sight extends a substantial 3' (12.8 kpc) above the midplane at its highest point.

\par
   The total mass we derive from our observations (Table~\ref{n4013tbl-1}) is close to that derived from single-dish data \citep{2005ApJS..160..149S}, which implies that there is no substantial flux missing.  A first assessment going by apparent extent only indicates that 32$\%$ of the total {\sc H\,i} mass is at a height greater than 1 kpc (14").  If only material within a radius of 7.9 kpc (2.2'), thus avoiding the clearly warped regions, this percentage amounts to 18$\%$.  Again, we emphasize that this is only the {\sc H\,i} with an apparent extent above 1 kpc.  Naturally if a line of sight warp component is present, much of this gas will be closer to the midplane.

\par
   We now consider tilted-ring models created to fit these data.

\section{The Models}\label{models}

\par
   For tilted-ring modeling, in which we create model galaxies comprised of a series of concentric rings for which parameters such as the rotational velocity, inclination, position angle, etc. may be specified, we use the Tilted Ring Fitting Code (\textrm{TiRiFiC}, \citealt{2007A&A...468..731J}).  Through tilted-ring modeling, features such as warps (via changes in inclination and position angle) and flares (vial radial increases in scale height) may be constrained. In addition to capabilities included in previous tilted-ring modeling software, \textrm{TiRiFiC} allows for the fitting of asymmetries and localized features, as well as lags, which we utilize in our models of NGC 4013.

\par
\begin{figure}
\centering
\includegraphics[width = 80mm]{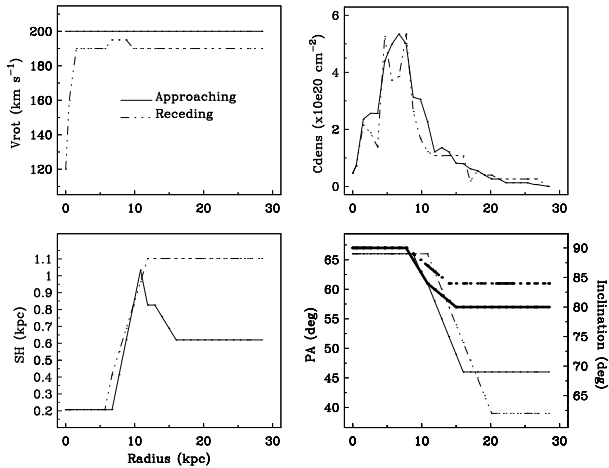}
\caption[Model Parameters for NGC 4013]{\scriptsize\textit{The parameters used in the final models of NGC 4013 for the approaching and receding halves.  Parameters presented here are consistent for all models except for the scale height and inclination distribution in the case of the 1C model, and the scale height distribution for the W model, for which these parameters are held constant by definition.  Note the relatively symmetric distributions of each quantity.  The bold line in the bottom-right panel represents the inclination} \label{4013multiplot}}
\end{figure}

    The approaching and receding halves of NGC 4013 are modeled separately as it is clear upon initial inspection that their morphologies differ greatly, and fitting them together could lead to unnecessary complexity in the models.  Additional asymmetries almost certainly exist, but due to projection effects, these are not obvious upon inspection.  Thus, we only allow for asymmetry between the two halves.  The initial estimate for the inclination was 90$^{\circ}$ and the position angle (PA) for the non-warped disk is 65$^{\circ}$, both determined by eye (with only very slight modifications to the central PA during later stages of the modeling and no modification to the central inclination).  The initial rotation curve was estimated based on matching the velocities on the terminal sides of the data and model lv (major-axis position-velocity) diagrams.  Subsequent modifications to the rotation curve were made in the same manner until optimal fits were reached.  The initial surface brightness distribution was set by examining the flux distribution along the midplane by eye.  The surface brightness distribution was altered only slightly during the modeling, with the final distribution optimized for our final preferred model, and imposed on all other models.  The reasoning behind imposing the surface brightness distribution from the final model onto the other models is 1) any differences between models were minute and could be attributed to degeneracies between the rotation curve and surface brightness, and 2) to maintain a clear and consistent picture of the changes in the models due to each additional component. All models presented here share the same surface brightness distribution, velocity dispersion (15 km s$^{-1}$), systemic velocity (835 km s$^{-1}$), central position (11h 58m 31.6s, 43$^{\circ}$ 56' 52.2"), central inclination (90$^{\circ}$), radial position angle dependence, and rotation curve.  The only exception is that the models including a lag have a rotation curve that is 10 km s$^{-1}$ higher in the approaching half than those that do not.  The parameters not given here are shown in Figure~\ref{4013multiplot} for final models.

\subsection{Individual Models}

\par
   We now address individual models and their defining characteristics, starting with a simple one-component model with a warp component across the line of sight (1C in figures).

\par
\begin{figure}
\centering
\includegraphics[width = 80mm]{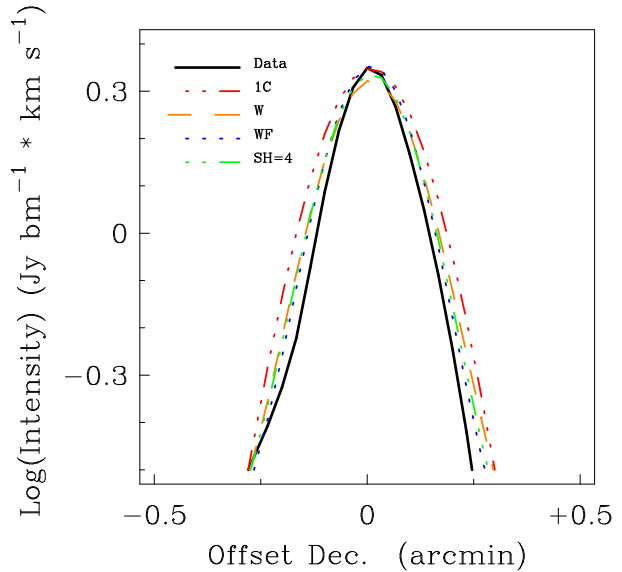}
\caption[Vertical Profiles of NGC 4013 and Models]{\scriptsize\textit{Vertical profiles of the data, one-component (1C), line of sight warp (W), and warp with a flare (WF), summed over a range of $\pm$60" (4.3 kpc) from the center in order to avoid as much of the warped regions as possible.  Additionally, we present the WF model with a central exponential scale height of 4" as opposed to 3" in order to illustrate the subtle difference between the two since such differences are pushing the limits of the resolution of the data.  Note how the 4" model slightly underestimates the peak.  Given this very slight preference, we use the 3" scale height in our final models, although the central scale height could indeed be closer to 4".  A color figure is available in the online version.} \label{4013vprofsumlog}}
\end{figure}
\begin{figure*}
\centering
\includegraphics[width = 160mm]{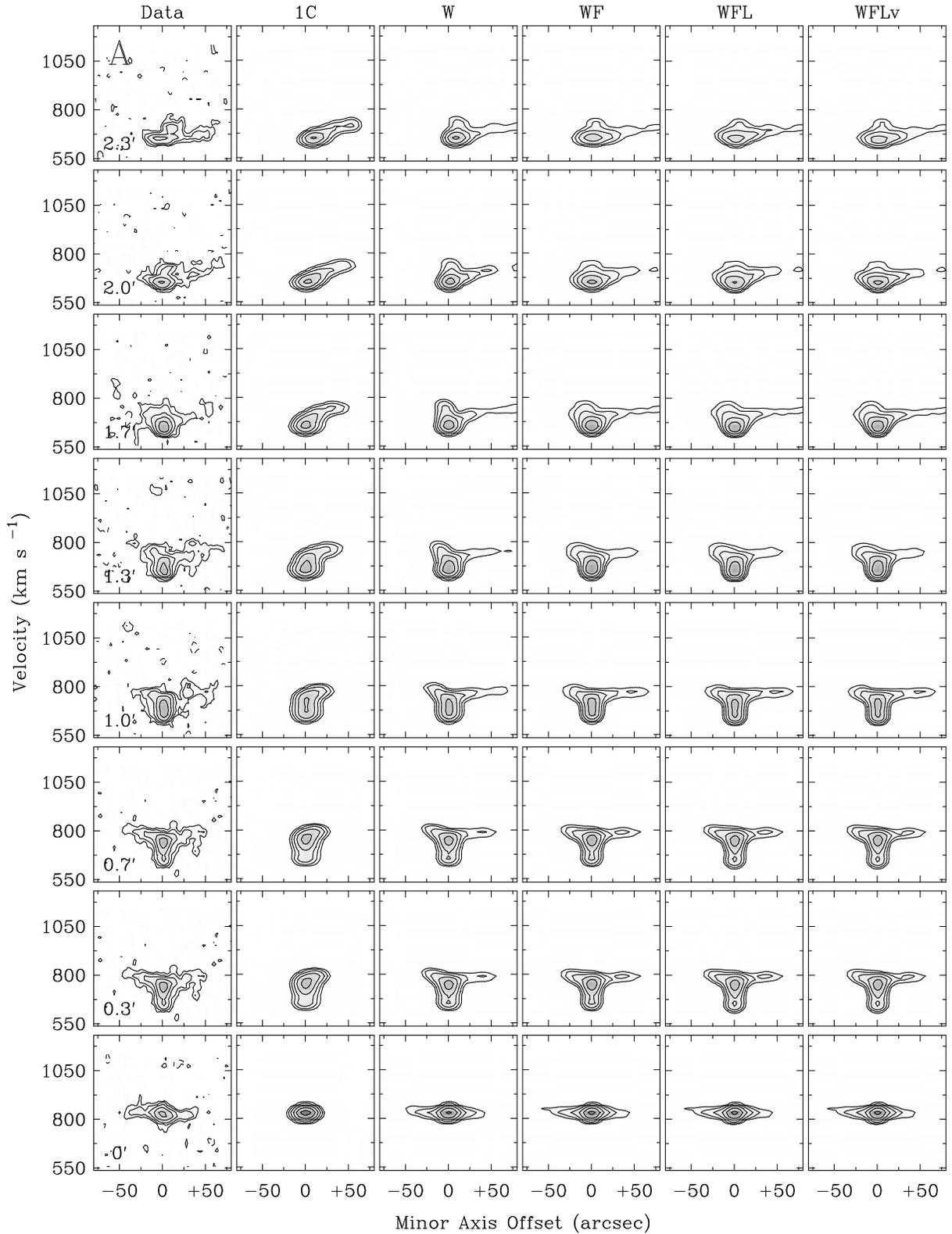}
\caption[Minor Axis Position-velocity Diagrams of NGC 4013 and Models]{\scriptsize\textit{bv diagrams showing all of the models presented in this paper.  Note the clearly poor fit of the 1C model, especially in the narrow systemic and wide terminal sides of each panel, as well as near the center.  The W model is an improvement, but still lacks the proper distribution near the systemic sides that is achieved through the addition of a flare. Contours are as in Figure~\ref{channelmapsymmetric4013}.  Note the improved match in the curvature on the terminal sides of panels corresponding to $\pm$2.0' and $\pm$2.3' that is due to the lag in this model.} \label{4013bvapproaching}}
\end{figure*}
\begin{figure}
\centering
\includegraphics[width = 50mm, angle = 270]{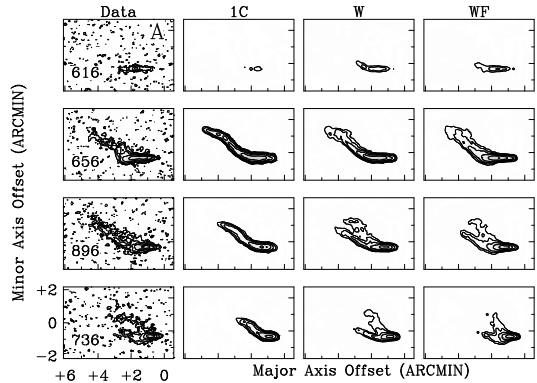}
\caption{\scriptsize\textit{Representative channel maps showing the approaching (A) and receding (B) halves of NGC 4013.  Note the splitting in the panel corresponding to 736 km s$^{-1}$ and how this is achieved by adding a warp component along the line of sight.  The 1C, W, and WF models are shown here as these are the models exhibiting the most substantial changes from one another.  The WFL and WFLv models are omitted from this figure as they are very similar to the WF model in channel maps since the lag is only in the flat part of the disk, which is visually overwhelmed by the outer, warped region here.   Contours are as in Figure~\ref{channelmapsymmetric4013}} \label{4013channelmapapproaching}}
\end{figure}

  As stated previously, it is clear that there exists a warp component in NGC 4013 that lies across the line of sight.  Thus, we model it with a single disk having an exponential scale height optimally found to be 7"$\pm$0.5 (500$\pm$35 pc).  To this model we add a PA warp for which parameters are given in Figure~\ref{4013multiplot}.  This model fails to match the vertical profile [summed over $\pm$60" (4.3 kpc) from the center in order to avoid the warped outer regions as much as possible (Figure~\ref{4013vprofsumlog})].  The width of the vertical profile is overestimated, yet the model cannot reproduce the spread near the systemic side of bv (minor-axis position-velocity) diagrams (Figure~\ref{4013bvapproaching}).  Thus, the model is too wide on the terminal side, but too narrow on the systemic.   Similarly, as may be seen in channel maps shown in Figure~\ref{4013channelmapapproaching}, the disk is too thick at small radii, but too narrow at large radii.  The splitting of most of the data bv diagrams near the systemic side is indicative of a warp component along the line of sight, which we now address.

\begin{figure*}
\centering
\includegraphics[width = 160mm]{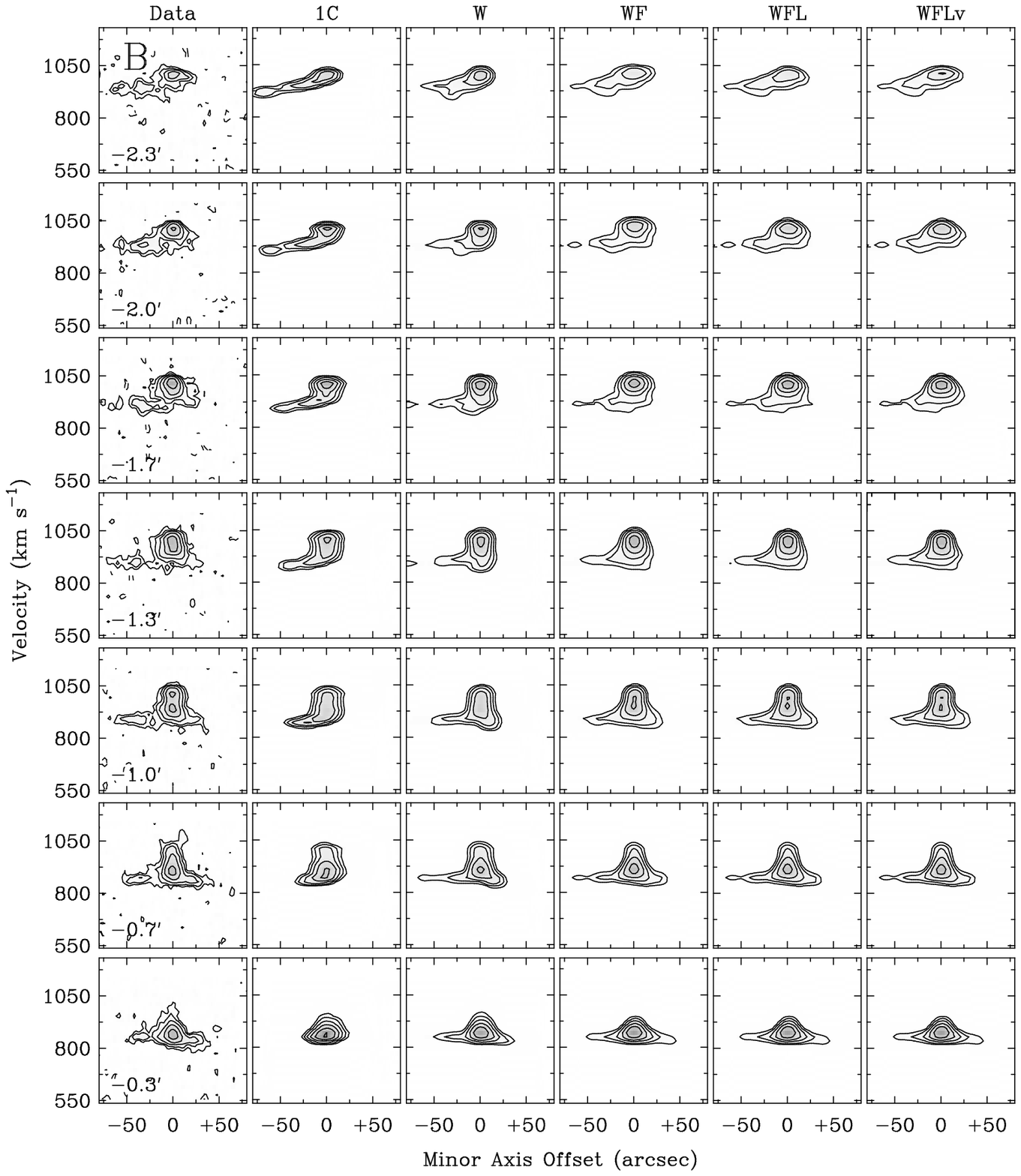}
\end{figure*}

\begin{figure}
\centering
\includegraphics[width = 50mm, angle = 270]{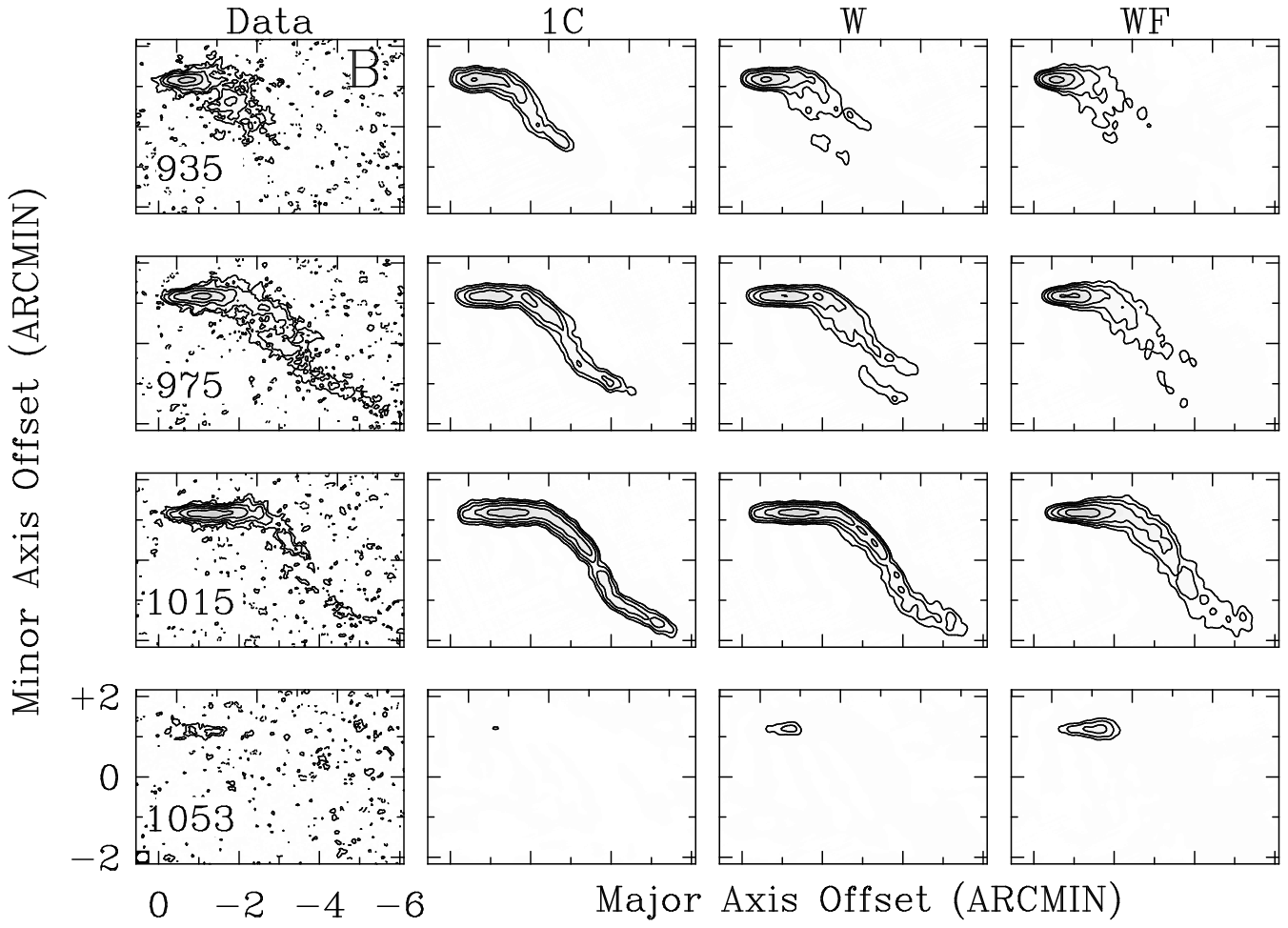}
\end{figure}

\par
   We add a line of sight warp component (``W" model in figures, inclination parameters given in Figure~\ref{4013multiplot}) that begins near the radial onset of the other warp component ($\sim$1.2' or 10 kpc), but is roughly a quarter the magnitude.  The relative contributions of each component to the total warp indicate that, in reality, the warp is oriented at an angle approximately 70 $^{\circ}$ from the line of sight (of course, there is a degeneracy due to projection effects in terms of the near and far sides).  The scale height of the layer is reduced to 5" (350 pc) to accommodate for the effects of the warp in this model.  One can see clear improvements to the bv diagrams and channel maps, but can also see that the emission near the systemic side is not entirely matched in the former, nor are the outer edges of the latter.   The warp also improves the fit to the vertical profile, although a closer match is still desirable.  In the channel maps, the data show an increasing minor axis thickness with radius, especially in the unwarped component, which is not present in this model.  This suggests a flare.

\par
   Adding a substantial flare to the model (``WF" in figures) by increasing the scale height with radius allows for a better fit to the vertical profile (parameters shown in Figure~\ref{4013multiplot}).  Additionally, the scale height of the approaching half decreases slightly for radii larger than 3.1' (13.2 kpc).  The emission on the systemic sides of bv diagrams is now spread out more, and the aforementioned thickening in the channel maps is well reproduced.  The improvements due to the flare are best seen on the systemic sides (going from systemic to $\sim$75 km s$^{-1}$ above systemic) of bv diagrams corresponding to radii greater than 1.0' (4.4 kpc) as seen in Figure~\ref{4013bvapproaching}.  Note the increased spread between the contours, which is in better agreement with the data.  Also note the increase in the spread along the minor axis. One issue to note is that to fit this model, the central scale height is reduced to 3" (210 pc), pushing the limits of our ability to constrain this parameter given our resolution.  However, we do see an \textit{extremely} slight distinction between a scale height of 3" and 4", mainly near the peak, shown in Figure~\ref{4013vprofsumlog}.  In the end, the significance of this difference is debatable, but we use the 3" (210 pc) scale height for subsequent models.

\par
    For the sake of completeness, a second, thicker disk is briefly considered, but as one may have already assumed, this yields no improvement to the models and is quickly dismissed.  For instance, the thickening in the minor axis direction evident in the channel maps as discussed above is naturally well matched by a flare.  A thick disk would not reproduce such behavior and add unnecessary complexity to the models.

\section{The Addition of a Lag}
\par
\begin{figure}
\centering
\includegraphics[width = 80mm]{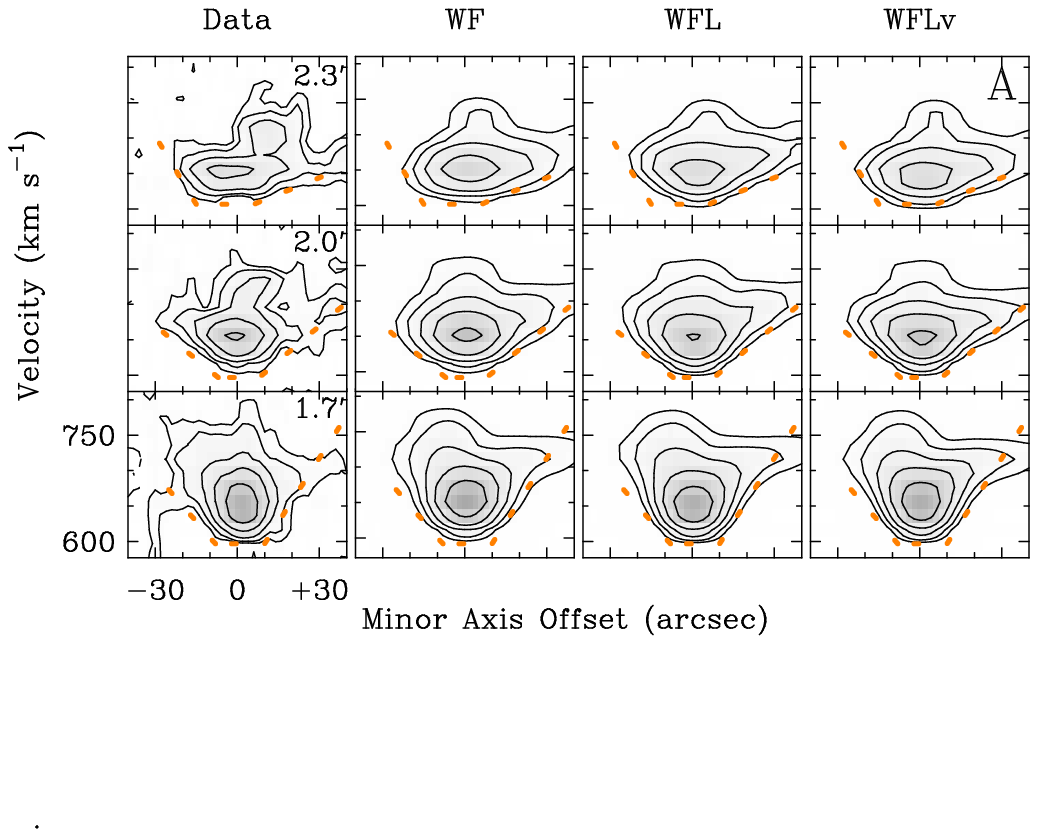}
\caption{\scriptsize\textit{Select bv diagrams of the data, warped model with a flare (WF), warped model with a flare and constant lag (WFL) , and warped model with a flare and radially shallowing lag (WFLv) for the approaching (A) and receding (B) halves.  The orange lines trace the approximate shape of the terminal sides of the bv diagrams in the data. In Figure A, note the overall rounded appearance of the WF model compared to the data.  Note also how the terminal side of the WFL model is too pointed in the panel corresponding to 2.3', while the WFLv is somewhat rounded, in closer agreement with the data.  In Figure B, the WF model is again too rounded compared to the data.  Regarding the radially shallowing lag, note the rounded appearance of the second and third contours on the terminal side of the data at $-$2.3'.  The WFL and WFLv models both fit these curves quite well at $-$2.3', with the WFL model perhaps being slightly too pointed.   However, at $-$2.0' and $-$1.7', it becomes evident that the lag in the WFL model is too shallow, especially at larger minor axis offsets, closer to 950 km s$^{-1}$. Thus, to fit the data at smaller radii, the lag would need to become steeper for the WFL model, resulting in too much of a V-shape in the $-$2.3' panel. Contours are as in Figure~\ref{channelmapsymmetric4013} A color figure is available in the online version.} \label{bvapproachinglag_zoom_lines}}
\end{figure}

     Improvements are seen via the addition of a lag of $-$21$^{+7}_{-14}$ km s$^{-1}$ kpc$^{-1}$ starting at a radius of 1.4' (5.8 kpc). The lag causes the observed velocity of the gas to be shifted in the direction of the systemic velocity, with this shift increasing with height.  This causes the terminal side of the contours in bv diagrams to shift from a relatively shallow U-shape to a steeper V-shape.  Although we attempt to alter as few parameters as possible, a slight adjustment to the rotation curve in the approaching half (190 km s$^{-1}$ to 200 km s$^{-1}$) yielded a better fit for the lag models. The subtle improvements from the addition of such a lag are best seen in panels corresponding to the terminal sides of $\pm$2.0'-2.3' of Figures~\ref{4013bvapproaching} and~\ref{bvapproachinglag_zoom_lines} (WFL). (Recall that there is a flaring layer in NGC 4013, so there is more {\sc H\,i} present at high z at large radii, making it easier to detect a lag.)  As for the central regions within $\sim$1.3', due to a lack of spatial resolution relative to the thickness of the disk, we cannot use the bv diagrams to meaningfully constrain the lag in those regions.

\par
\begin{figure}
\centering
\includegraphics[width = 80mm]{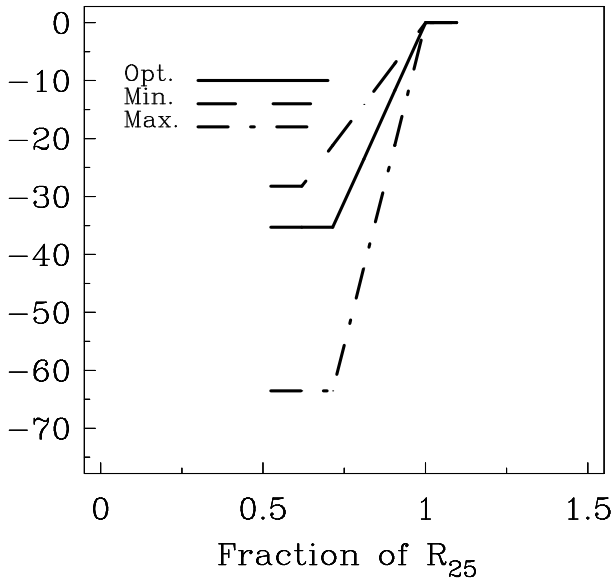}
\caption{\scriptsize\textit{The distributions of lag magnitudes for the minimum, optimal, and maximum lag models shown in Figure~\ref{bvapproachinglag_zoom}. Constant lag values are kept to as large of radii as possible before it is obvious that radially shallowing is necessary.  Thus, the lag may shallow over its entire range. Note the increased uncertainty in the upper limits due to the spatial resolution along the minor axis of the data leading to smaller changes introduced by steeper lags.} \label{parameters_lag}}
\end{figure}

    We see further improvement by allowing the lag to shallow with radius (WFLv) in Figures~\ref{4013bvapproaching} and~\ref{bvapproachinglag_zoom_lines}.  Note how the contours on the terminal side in the panels corresponding to $\pm$1.7' and $\pm$2.0' have a ``V" shape on the terminal side in both the data and radially varying lag models, while those at $\pm$2.3' are distinctly more rounded in both the data and the radially varying lag model than in the constant lag model.  This behavior can only be reproduced if the lag is allowed to shallow radially.  The shallowing itself is parametrized in Figure~\ref{parameters_lag}.

\begin{figure}
\centering
\includegraphics[width = 80mm]{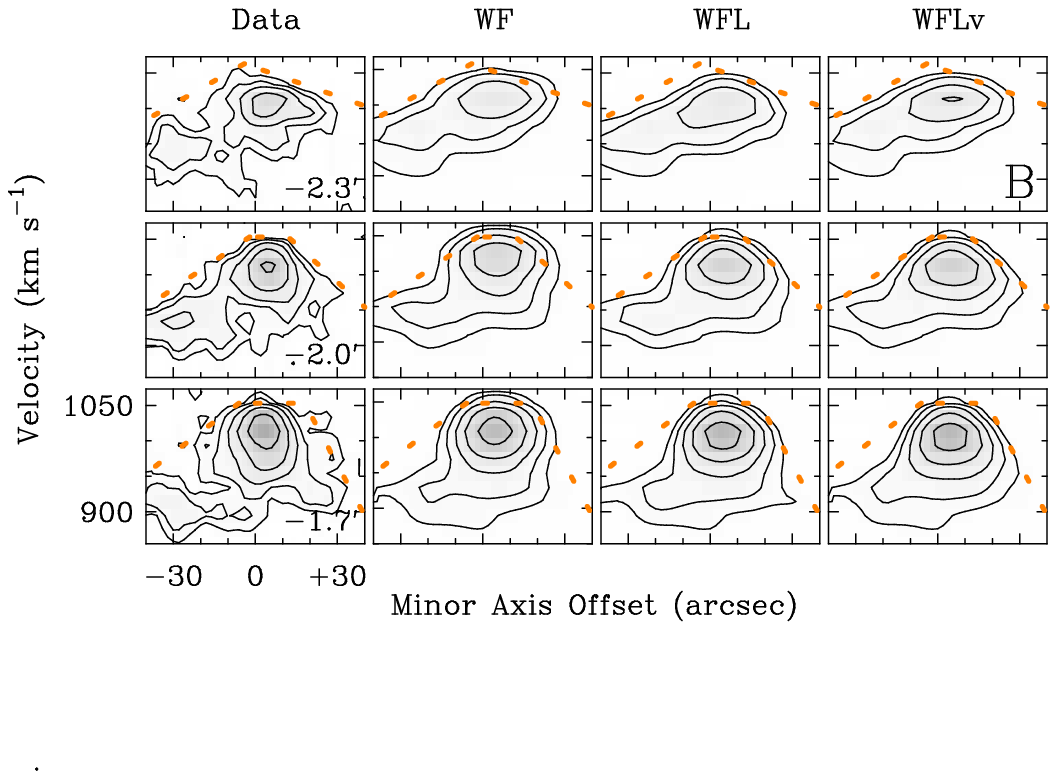}
\end{figure}

\par
\begin{figure}
\centering
\includegraphics[width = 80mm]{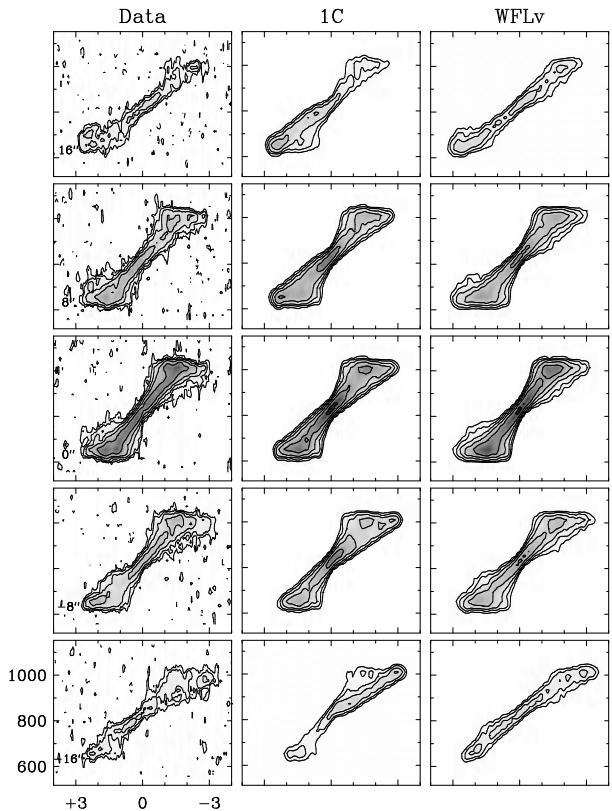}
\caption[Major Axis Position-velocity Diagrams of NGC 4013 and Models]{\scriptsize\textit{lv diagrams showing the data, one component (1C), and final (WFLv) models.  The most substantial changes are near the systemic side and at large minor axis offsets.  Contours are as in Figure~\ref{channelmapsymmetric4013}} \label{4013lvplotpaper}}
\end{figure}

   For completeness, we show lv diagrams of the 1C and final WFLv models in Figure~\ref{4013lvplotpaper}.

\section{The Uncertainties of the Models}
\par
    The rotation curves on each half nearly match each other, and the average uncertainty in individual rings is approximately 5 km s$^{-1}$.  The runs of the position angle and inclination are both extremely sensitive to small changes of a degree or two in a single ring.  The narrow scale height near the center should be considered somewhat skeptically as it approaches the limits of our ability to constrain this parameter given the resolution of the data and should be considered an upper limit.  Still, a slight change can be seen in the vertical profile when the scale height near the center is changed from 3" to 4" (Figure~\ref{4013vprofsumlog}).  It is also clear that there is a flaring layer in NGC 4013.

\begin{figure}
\centering
\includegraphics[width = 80mm]{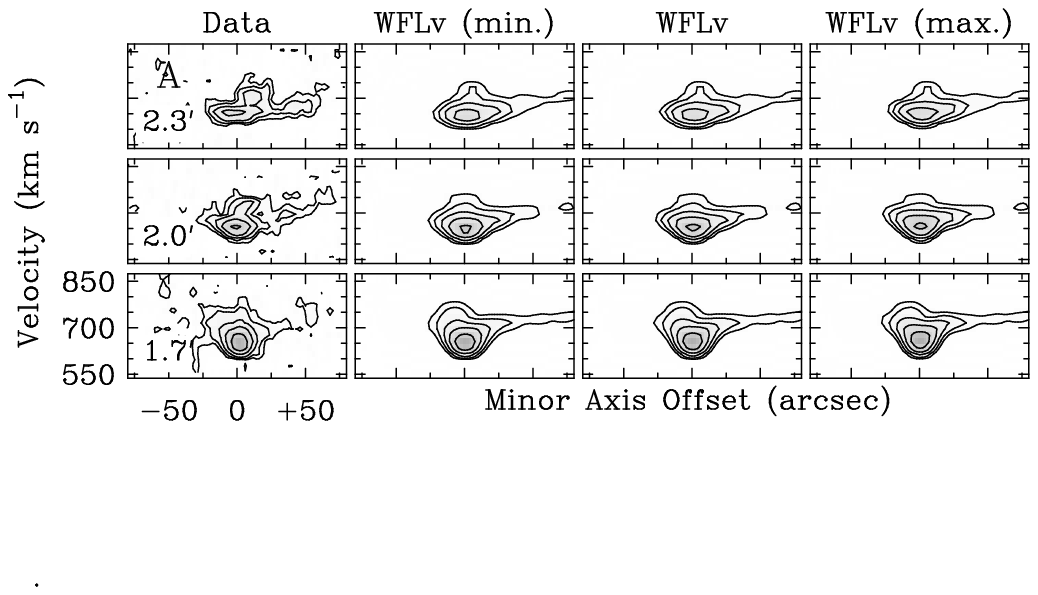}
\caption{\scriptsize\textit{Select bv diagrams of the data, minimum, optimal, and maximum lag models for the approaching (A) and receding (B) halves.  The various lags are parametrized in Figure~\ref{parameters_lag}.  Note in particular the subtle differences in the curvature along the terminal sides in each model.  The minimum lag model is too rounded, while the maximum model has a distinct ``V" shape not present in the data. Contours are as in Figure~\ref{channelmapsymmetric4013}} \label{bvapproachinglag_zoom}}
\end{figure}

\begin{figure}
\centering
\includegraphics[width = 80mm]{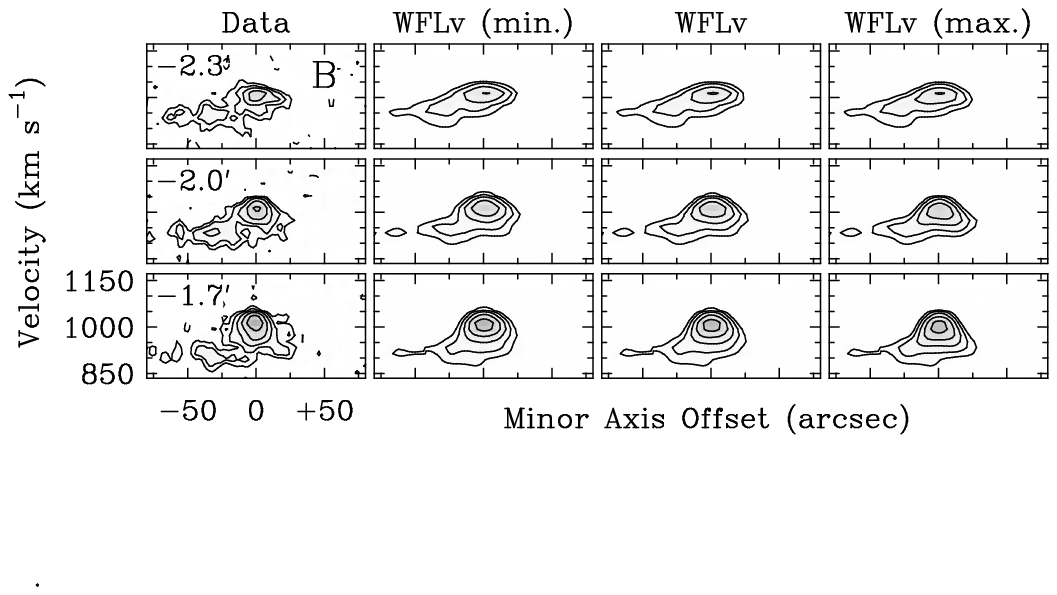}
\end{figure}

\par
    We consider uncertainties in the lag by creating minimum and maximum value models which are judged to be still consistent with the data in addition to the optimal lag (Figure~\ref{bvapproachinglag_zoom}).  The various lag distributions are parametrized in Figure~\ref{parameters_lag}.  Note the high maximum lag allowed, which is due to the thinness of the disk in relation to the resolution of the data.  This difficulty is because of the thinning along the minor axis of the terminal side of the bv diagrams when steeper lags are introduced.  The thickness on this side becomes insensitive to the lag for relatively steep lags.  For a similar reason, an upper limit for the magnitude of the lag is increasingly difficult to determine with decreasing radius.  Within $\sim$7 kpc, our small upper limit on the scale height limits our ability to constrain steeper lags that result in even further thinning along the terminal side of the minor axis.   

\par
    By summing the square of the residuals for each model interior to 1.15' (10.2 kpc), we attempt to quantify the improvement due to each feature.  We do not consider the addition of a line of sight warp component (W) to a (1C) model here since it is in reality connected to the warp component across the line of sight which is obvious upon inspection.   Adding a flare (WF) to the warped (W) model yields a decrease of 3.7\% in the approaching half and 2.6\% in the receding half.  The addition of a constant lag results in a total increase with respect to the W model of 11.3\% and a decrease of 4.9\%, respectively, while allowing the lag to shallow radially provides an decrease of 8.1\% in the approaching and a decrease of 5.3\% in the receding compared to the W model.   

\par
    The rounding on the terminal side of the bv diagram at 2.3' in the approaching half indicates some degree of shallowing, and yet the numerical goodness of fit is worse than for the model with a constant lag (for the approaching half only). This inconsistency highlights the problem inherent in tilted-ring modeling of galaxies, in particular their faint and often locally-disrupted extra-planar layers, which greatly hinders automated fitting efforts, rendering a by-eye approach necessary as demonstrated throughout the literature (e.g.\ Fraternali et al. 2004/2005, \citealt{2007AJ....134.1019O}, Zschaechner et al. 2011/2012/2015, \citealt{2013MNRAS.434.2069K}, \citealt{2013A&A...554A.125G}).  The effects of lags predominantly occur in the faintest gas - only a fraction of the total emission.  Thus, the quantifiable improvements from them are suppressed over large regions.  Furthermore, if bright material close to the midplane is displaced by a lag (as is the case here), the detriment from that dominates the final quantization of the overall improvement to the models.  This issue is particularly relevant to NGC 4013 given its thin disk, which pushes the limits of the resolution of these data.  However, resolution is not an issue for the fainter, more diffuse emission extending above the midplane in which the lag is clearly observed and there are indications of its shallowing.

\par
     What these models demonstrate is that the overall characteristic V-shape of the terminal sides of the bv diagrams, as seen in other galaxies with lagging extra-planar {sc H,\i} (e.g.\ NGC 891, NGC 4565, NGC 4244) \textit{can only be achieved through the addition of a lag}.  The presence of this V-shape is independent of the rotation curve at the midplane.  Furthermore, the rounding of this V-shape at large radii, also demonstrated in the aforementioned galaxies, is achieved through shallowing of the lag.  It is possible that the rounding which we attribute to shallowing here, could be indicative of a constant lag starting higher above the midplane with increasing radius (not shown).  However, such a scenario, in which the rounding of the terminal sides of bv diagrams is attributed solely to starting height of the lag, is unlikely, although it, or a combination of shallowing and variable starting height cannot be ruled out based on these data alone.

\section{Discussion \& Conclusions}\label{4013discussion}
\subsection{NGC 4013}

\par
    The central {\sc H\,i} layer in NGC 4013 is rather thin with an upper limit of 4" (280 pc) for the central scale height.  The layer then flares up to 15" (1 kpc) at radii greater than 7 kpc.  The rotation curve peaks at 195 km s$^{-1}$ and is approximately flat, indicating a large dynamical mass (calculated as 1.1$\times$10$^{12}$ M$_{\odot}$ using a rotational velocity of 190 km s$^{-1}$ at a radius of 13.3 kpc).  We see no evidence for substantial amounts of extra-planar {\sc H\,i}, but see evidence for a radially shallowing lag with a peak magnitude of$-$35$^{+7}_{-28}$ km s$^{-1}$ kpc$^{-1}$ where the {\sc H\,i} layer flares.  The lag then shallows with radius, going to zero near R$_{25}$.  The lag is modeled starting from the midplane.  

\par
   Our models agree remarkably well with those presented in \citet{1996A&A...306..345B}, the most notable difference being a 20-30 km s$^{-1}$ decrease in rotational velocity in the outer warped radii in the \citet{1996A&A...306..345B} model, while our rotation curve remains flat.  This difference is likely due to degeneracies in modeling substantially warped regions.  Recall that the lag is considered only at radii interior to the warp, thus this discrepancy has no bearing on the final conclusions involving the lag.

\par
   The lack of extra-planar {\sc H\,i} is intriguing given the extra-planar DIG and dust.  However, it is possible that there is a relatively large escape of ionizing radiation into the gaseous halo, resulting in prominent extra-planar DIG but not {\sc H\,i}.

\par
   As a side note, we see no evidence for the potential {\sc H\,i} clouds noted by \citet{1995A&A...295..605B} that are slightly beyond and near the highest point in the warp on the approaching half and conclude (as \citet{1995A&A...295..605B} suggested) that they are likely artifacts in those data.

\par
    As is also noted by \citet{1995A&A...295..605B}, there are no indications of a bar in {\sc H\,i}, although one appears to exist in CO and at optical wavelengths \citep{1999A&A...343..740G}.  The location of the bar they describe coincides with the central region of lower column densities before the initial peak in {\sc H\,i} (Figure~\ref{4013multiplot}).


\subsection{The Lag in NGC 4013 in the Context of Other Galaxies}

\par
\begin{figure}
\centering
\includegraphics[width = 80mm]{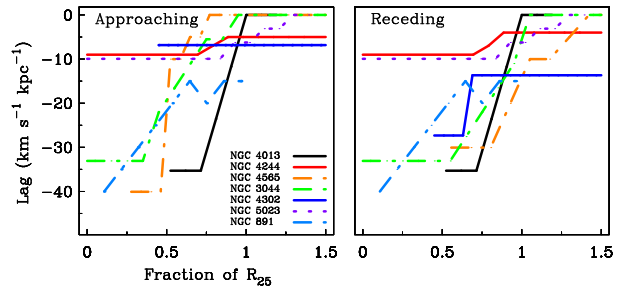}
\caption[]{\scriptsize\textit{Measured radial variations of lags in several nearby edge-on spiral galaxies including NGC 891 \citep{2007AJ....134.1019O}, NGC 4244 \citep{2011ApJ...740...35Z}, NGC 4565 \citep{2012ApJ...760...37Z}, NGC 5023 \citep{2013MNRAS.434.2069K}, NGC 4302 and NGC 3044 \citep{2015ApJ...799...61Z}.  Note the general trend of lags beginning to shallow near 0.5R$_{25}$, and reaching their shallowest values near R$_{25}$ itself. A color figure is available in the online version.} \label{parameterslag_new}}
\end{figure}

   The lag in NGC 4013 is the steepest measured for {\sc H\,i} to date.  It also shows radial shallowing similar to several existing measured lags (Figure~\ref{parameterslag_new}).  For a more extensive presentation involving lag properties among a larger sample of galaxies see \citet{2015ApJ...799...61Z}, in particular Table 5.  The addition of NGC 4013 does not change any conclusions concerning lags and SF properties already presented in that work.

\subsection{Implications for Theoretical Scenarios}\label{theory}
\par
   As has been previously noted, observed lag magnitudes are substantially steeper than those predicted by purely ballistic models (e.g.\ \citealt{2002ApJ...578...98C}).  The lag in NGC 4013 is no exception.  Recent models by \citep{2011MNRAS.415.1534M} consider a momentum exchange between galactic fountain clouds and an initially static, but subsequently slowly-rotating hot halo that is built up via accretion.  As noted by \citep{2015ApJ...799...61Z}, the radial shallowing observed in these galaxies indicates that if lags are indeed due to such a scenario, then this momentum exchange must be occurring predominantly within R$_{25}$.  

\par
   The shallowing itself is not yet understood, and must be considered in future simulations.  Benjamin (2002, 2012) suggests that extra-planar pressure gradients could impact lags and their radial variations.  There have been few observational constraints to date for such pressure gradients, a situation soon to be remedied by deep continuum observations of edge-on galaxies \citep{2012AJ....144...43I}.  Their sample includes seven galaxies with measured lags, including five (as well as the receding side of NGC 4302) with radially shallowing lags.


\section{Acknowledgments}
     We thank the operators at VLA for overseeing the observations.  The National Radio Astronomy Observatory is a facility of the National Science Foundation operated under cooperative agreement by Associated Universities, Inc.  This research has made use of the NASA/IPAC Infrared Science Archive, which is operated by the Jet Propulsion Laboratory, California Institute of Technology, under contract with the National Aeronautics and Space Administration.  We also thank Rene A. M. Walterbos, Gregory B. Taylor and Peter Zimmer for constructive comments.  This material is based on work partially supported by the National Science Foundation under grant AST-0908106 to R.J.R.  We thank the anonymous referee for exceptionally insightful and constructive comments regarding the presentation of these models.

\bibliographystyle{apj}
\bibliography{n4013.bib}

\end{document}